\documentclass[conference]{IEEEtran}
\IEEEoverridecommandlockouts
\usepackage{cite}
\usepackage{amsmath,amssymb,amsfonts}
\usepackage{algorithmic}
\usepackage{graphicx}
\usepackage{textcomp}
\usepackage{xcolor}
\usepackage{booktabs}
\def\BibTeX{{\rm B\kern-.05em{\sc i\kern-.025em b}\kern-.08em
    T\kern-.1667em\lower.7ex\hbox{E}\kern-.125emX}}
\begin{document}

\title{Predicting ulcer in H\&E images of inflammatory bowel disease using domain-knowledge-driven graph neural network%
\thanks{$^1$ Work done during an internship at Merck; corresponding author is Lin Li: lin.li23@merck.com}
}

\author{
\IEEEauthorblockN{
\parbox{\textwidth}{
\centering
Ruiwen Ding$^1$ ~~ Lin Li$^2$ ~~ Rajath Soans$^2$ ~~ Tosha Shah$^2$ ~~ Radha Krishnan$^2$ ~~\\
Marc Alexander Sze$^2$ ~~ Sasha Lukyanov$^2$ ~~ Yash Deshpande$^2$ ~~ Antong Chen$^2$
}
}
\\[1ex]
\IEEEauthorblockA{$^1$ University of California, Los Angeles, CA, USA}
\IEEEauthorblockA{$^2$ Merck \& Co., Inc., Rahway, NJ, USA}
}

\maketitle

\begin{abstract}
Inflammatory bowel disease (IBD) involves chronic inflammation of the digestive tract, with treatment options often burdened by adverse effects. Identifying biomarkers for personalized treatment is crucial. While immune cells play a key role in IBD, accurately identifying ulcer regions in whole slide images (WSIs) is essential for characterizing these cells and exploring potential therapeutics. Multiple instance learning (MIL) approaches have advanced WSI analysis but they lack spatial context awareness. In this work, we propose a weakly-supervised model called DomainGCN that employs a graph convolution neural network (GCN) and incorporates domain-specific knowledge of ulcer features, specifically, the presence of epithelium, lymphocytes, and debris for WSI-level ulcer prediction in IBD. We demonstrate that DomainGCN outperforms various state-of-the-art (SOTA) MIL methods and show the added value of domain knowledge.
\end{abstract}

\begin{IEEEkeywords}
Graph neural network, digital pathology, weakly-supervised learning, inflammatory bowel disease
\end{IEEEkeywords}

\begin{figure*}[]
\centering
\includegraphics[width=0.9\linewidth]{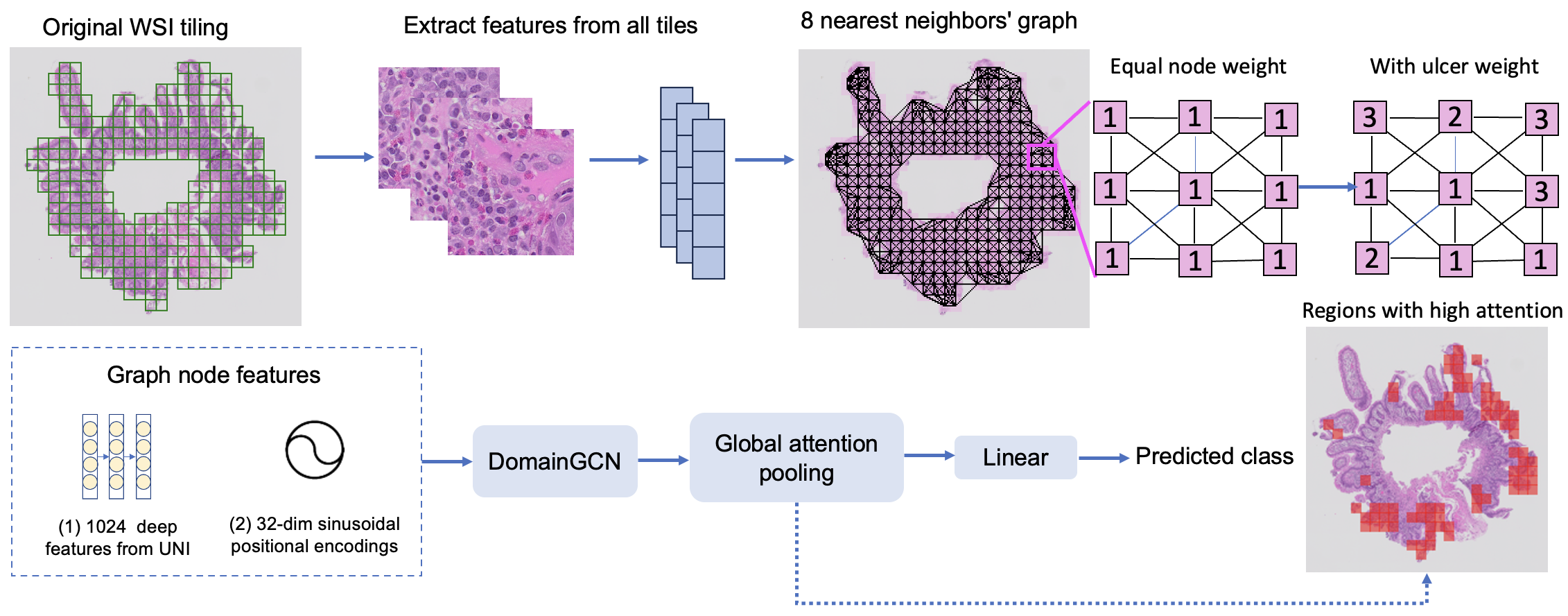}
\caption{The pipeline of DomainGCN including WSI patching, node feature extraction, 8-nearest-neighbor graph construction, and graph modeling.}
 \label{fig:pipeline}
\end{figure*}

\section{Introduction}
Inflammatory bowel disease (IBD) refers to a group of chronic inflammatory conditions that primarily affect the digestive tract. It is characterized by the complex interplay of several factors such as genetics, microbiome, immunity, and environmental triggers \cite{baumgart2007inflammatory}. The intestinal immune cells, e.g. neutrophils, macrophages, and lymphocytes, significantly influence the immunological responses in IBD \cite{lv2024comprehensive}. Therefore, studying the immune cell characteristics can improve our understanding of IBD development and help investigate potential therapeutic methods. Disease activity as measured histologically is ascertained by the presence of neutrophilic granulocytes and their damage to the epithelium \cite{seldenrijk1991histopathological}. Ulcer regions in IBD tissue samples contain rich inflammation information, such as neutrophilic inflammation, which is rarely found in normal regions\cite{vespa2022histological}. Therefore, accurate identification of ulcer regions can provide regions of interest for further immune cell identification and characterization.  

 Fully-supervised models can be trained to automatically segment the ulcer regions but requiring laborious and time-consuming spatial annotations. Weakly-supervised deep learning approaches, such as multiple instance learning (MIL), address the challenge of lacking spatial annotations by using whole slide image (WSI) level labels\cite{lu2021data, shao2021transmil}. The pipeline usually involves breaking the WSI down into small non-overlapping patches and aggregating patch features using attention scores to generate a WSI-level prediction \cite{lu2021data}. However, MIL models assume patch independence without capturing spatial relationships. To address this issue, some studies have used graph neural network (GNN)-based approaches, such as graph convolution neural networks (GCNs), to capture spatial connectivity between patches in WSIs by taking each WSI as a graph with the image patches as the nodes to capture the spatial connectivity between nodes \cite{ding2022spatially, chen2021whole}. However, most existing works in GNNs do not explicitly incorporate any domain-specific information in graph modeling. 
 
 Ulcer regions in IBD are usually characterized by multiple local architectural features such as detachment of epithelium from the basal membrane, lymphocyte infiltration, bleeding, and presence of debris \cite{villanacci2021histopathology}. Since GNNs are designed to capture the local spatial context of graphs, they likely will be more effective at capturing ulcer regions as compared to MIL-based approaches which are not spatial-context-aware. In addition, during the message-passing operation, it might be beneficial to incorporate domain knowledge by placing more importance on the nodes that exhibit more prominent ulcer features.

In this work, we devise a GCN-based weakly supervised model DomainGCN to classify whether there is an ulcer in a hematoxylin and eosin (H\&E)-stained WSI of IBD. Specifically, we incorporate ulcer domain knowledge related to the abundance of epithelium, lymphocytes, and debris during the message passing operations of GCN. We hypothesize that DomainGCN can be more accurate at WSI-level classification as compared to the non-GNN-based MIL methods. The contributions of our work are as follows:

1. We incorporate domain knowledge related to ulcer features in our GNN modeling and showed its added value.

2.  We show that in general, GNNs are more accurate at the WSI-level ulcer prediction as compared to MIL methods. 

3. To the best of our knowledge, this is the first work exploring the effectiveness of GNNs in ulcer predication and localization in  H\&E-stained WSIs of IBD patients.

\section{Methods}

\subsection{Dataset}
\label{dataset}
A total of 667 WSIs of biopsy tissues scanned at 40x magnification were available at the time of analysis from IBD Plexus database with informed consent from Crohn’s and Colitis Foundation \cite{raffals2022development}. Each patient has only one WSI. We included patients whose biopsy sample was taken from the ileum since we wanted to control for any morphological heterogeneity associated with samples from different body locations. WSIs with substantial artifacts such as bubbles and corrupted WSI images were excluded. The WSI labels were originally provided as macroscopic appearance from the IBD Plexus database but upon re-review of these WSIs with our pathologist, we found that some labels were not consistent with the histopathological assessment and corrected for them. At the end, there were 305 WSIs used in this study with 183 (60\%) non-ulcer WSIs and 122 (40\%) ulcer WSIs. 

\subsection{Graph construction and model architecture}
\label{architecture}
Figure \ref{fig:pipeline} shows the pipeline of this work. Firstly, each WSI was patched into non-overlapping patches of size 512 by 512 at 40x, and then resized to size 256 by 256. Then two groups of features were extracted from each patch and were used as the initial input node features to the graphs. The first group is 1024 deep features extracted from a self-supervised model for digital pathology called UNI \cite{chen2024towards}. The second group is a 32-dimensional non-trainable sinusoidal positional encoding features \cite{dosovitskiy2020image} derived from the relative spatial coordinates of each patch within the WSI. The next step involves building a graph from each WSI, where patches are the nodes and edge connectivity is defined by k-nearest neighbors. In this work, k = 8 and Euclidean distance between patchs was used as the distance metric. The assumption is that the immediate neighboring patches provide spatial context for each other. 

The next steps involve initializing the graphs with the two groups of node features and applying the message-passing operation to learn spatial context-aware feature representations from the graphs. Suppose each WSI is represented by a graph and each graph has $N$ nodes. The two groups of node features are UNI node features $\boldsymbol{X_{UNI}}$ $\in$ $\mathbb{R}^{N \times 1024}$ and positional encoding node features $\boldsymbol{X_{PE}}$ $\in$ $\mathbb{R}^{N \times 32}$.  First, a linear layer was applied to transform $\boldsymbol{X_{UNI}}$ from 1024 dimensions to 64 hidden dimensions. Then $\boldsymbol{\hat{X}_{UNI}}$ was concatenated with $\boldsymbol{X_{PE}}$ to form the final node features $\boldsymbol{X}$ $\in$ $\mathbb{R}^{N \times 96}$. 

The architecture of DomainGCN was based on the one of GCN \cite{li2020deepergcn}. Generally, the message-passing GNNs work by iteratively aggregating information from the neighboring nodes and updating the current node features. After several iterations of such information update (one iteration is achieved by having one message-passing layer), the learned representation can reflect the topological structure of the graph data. GCN's message-passing approach is based on a softmax aggregation of neighboring node features and residual connections. Specifically, the message-passing function is:

$$
x_i^{l+1} = x_i^l + MLP(x_i^l + Softmax(ReLU(x_j^l) + \epsilon:j \in \mathcal{N}_i))
$$

where $x_i^l$ is the node's features in the $l$-th layer, $x_j^l$ is a neighboring node of $x_i^l$, $\epsilon$ is a small positive constant, and $x_i^{l+1}$ is the resulting updated node's features in the $(l+1)$-th layer after the message passing operation. Afterward, a global attention pooling layer was applied to aggregate the GCN-learned representations from all nodes and generate a WSI-level embedding. Once the model is trained, the attention scores from this global attention pooling layer were used to visualize the most-attended regions on the WSIs. After getting the WSI-level embedding, a linear layer was then applied to render the WSI-level prediction, which was optimized using cross-entropy loss with the true ulcer label.

\subsection{Incorporation of domain knowledge (ulcer weight)}
Before the message-passing operation, each node was weighted differently according to the amount of epithelium, lymphocytes, and debris present in each node/patch. Typically, the ulcer regions have less epithelium, more lymphocytes, and more debris as compared to non-ulcer regions. A TIAToolbox's ResNet patch-level tissue classifier pre-trained on 100,000 histological human colorectal cancer and healthy tissue was applied to each patch in our WSIs \cite{Pocock2022}. There were 9 classes available from their model but only epithelium, lymphocytes, and debris were the classes of interests in this project. Then the probabilities of a patch having epithelium, lymphocytes, and debris were respectively used as a surrogate measure of the amount of each tissue. Specifically, initially all nodes/patches have a weight of 1. Then the weight is increased by 1 if a node's epithelium probability is less than 0.1, increased by 1 if a node's lymphocyte probability is greater than 0.3, and increased by 1 if a node's debris probability is greater than 0.3 based on empirical observations. At the end, a node can have a weight ranging from 1 to 4, depending on the probability of the three types of tissue classes. 

\begin{figure*}[]
\centering
\includegraphics[width=0.7\linewidth]{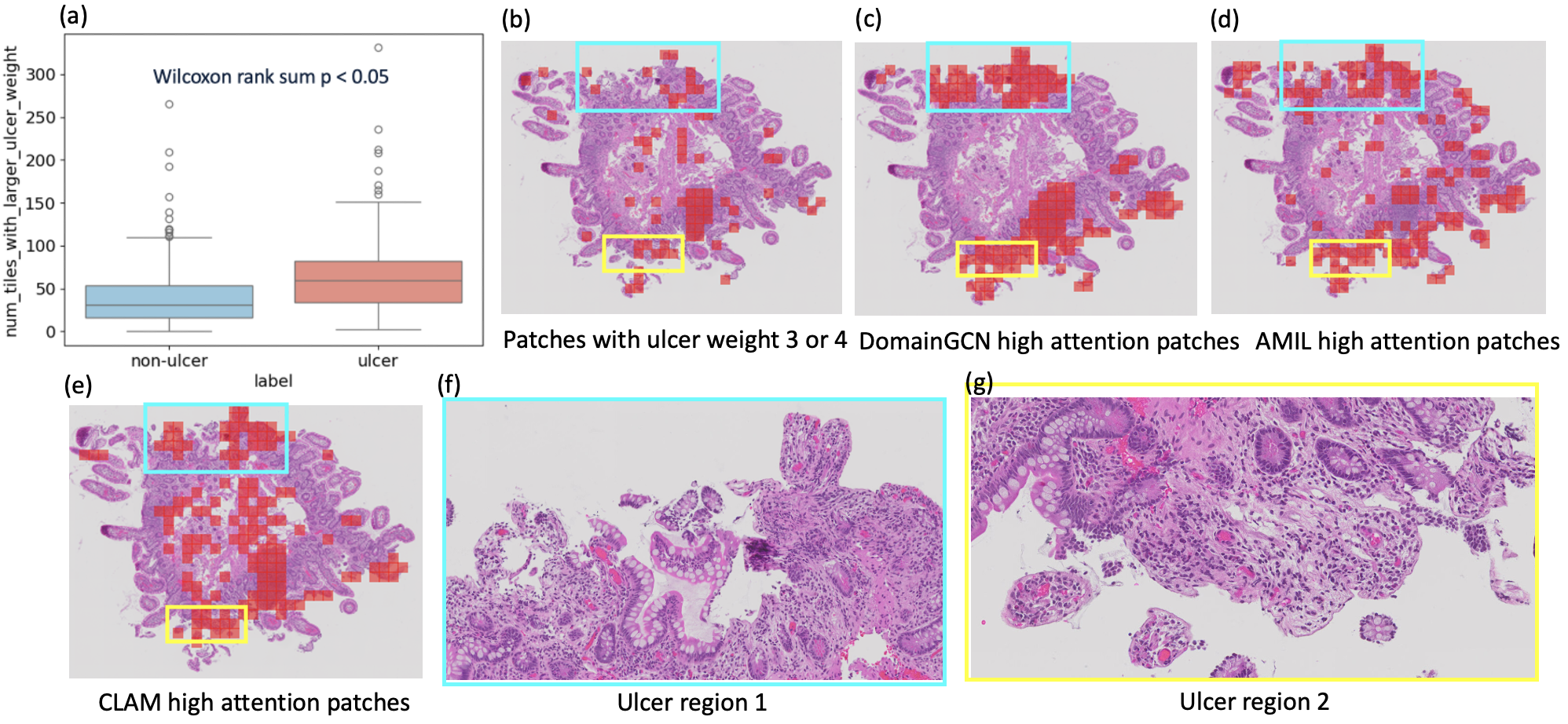}
\caption{Visualization of domain knowledge and attention maps. (a) represents boxplots showing the distribution of ulcer weights for non-ulcer and ulcer cases. (b) visualizes the patches that have ulcer weight 3 or 4. (c)(d)(e) respectively shows the patches with the top 25\% attention scores for DomainGCN, AMIl, and CLAM. (f) and (g) show the two ulcer regions of this case.}
 \label{fig:visualize}
\end{figure*}

Figure \ref{fig:visualize}(a) shows that the number of nodes/patches that have a relatively large ulcer weight of 3 or 4 for the ulcer-labeled WSIs was statistically significantly higher than the one of non-ulcer-labeled WSIs. Figure \ref{fig:visualize}(b) visualizes patches that have an ulcer weight of 3 or 4 on an example WSI.

\subsection{Baseline models}
The effectiveness of our model DomainGCN was compared against several SOTA MIL models and the graph models without domain information. All models' hyperparameters were tuned on our dataset with the original works' recommended parameters as a reference.

1) AttentionMIL: A MIL-based model that aggregates the patch-level features using attention scores learned by the model \cite{ilse2018attention}. 

2) TransMIL: A transformer-based MIL model that uses the Nyström method to approximate the importance of each patch during patch-level feature aggregation. \cite{shao2021transmil}

3) CLAM: An improved attention-MIL method that utilizes instance-level clustering over the highly attended patches to constrain and refine the learned feature space \cite{lu2021data}. CLAM with a single attention branch was used in this study.

4) MambaMIL: A Mamba-based MIL model that uses a state space model Mamba to calculate the importance of each patch during the patch-level feature aggregation. \cite{yang2024mambamil}

5) GPT: A graph transformer model \cite{zheng2022graph} that first passes the WSI-built graph into a GNN followed by a vision transformer to classify WSIs.

6) PatchGCN: A graph-based approach that uses GCN \cite{li2020deepergcn} as the GNN and builds graphs from WSIs using the 8-nearest neighbor approach to predict patient survival. 

7) GCN: The ablated version of DomainGCN where no ulcer weight was added to the model.

\begin{table}[]
\caption{Mean validation AUC, F1 score, and accuracy for all models. SD: standard deviation}
\vspace{0.2cm}
\centering{
\scalebox{0.75}{
\begin{tabular}{@{}lllll@{}}
\toprule
Models                   & AUC (SD) & F1 (SD) & ACC (SD)            \\ \midrule
AMIL \cite{ilse2018attention}           & 0.763 (0.0218) & 0.761 (0.0291) & 0.761 (0.0305)           \\
TransMIL \cite{shao2021transmil}  & 0.775 (0.0472) & 0.786 (0.0479) & 0.787 (0.0485)     \\
CLAM \cite{lu2021data} & 0.777 (0.0349) & 0.783 (0.0320) & 0.784 (0.0319)      \\
MambaMIL  \cite{yang2024mambamil}     & 0.781 (0.0284) & 0.778 (0.0282) & 0.777 (0.0286)   \\
GTP  \cite{zheng2022graph}   & 0.786 (0.0506) & 0.789 (0.0495) & 0.790 (0.0480)     \\
PatchGCN  \cite{li2020deepergcn}       & 0.790 (0.0516) & 0.787 (0.0535) & 0.787 (0.0529)   \\
GCN       & 0.792 (0.0193) & 0.783 (0.0260) & 0.784 (0.0262)  \\
DomainGCN         & \textbf{0.800 (0.00551)} & \textbf{0.794 (0.00826)} & \textbf{0.793 (0.00784)}  \\ \bottomrule
\end{tabular}
}
}
\label{table: table1}
\end{table}

\begin{figure}[]
\centering
\includegraphics[width=0.6\linewidth]{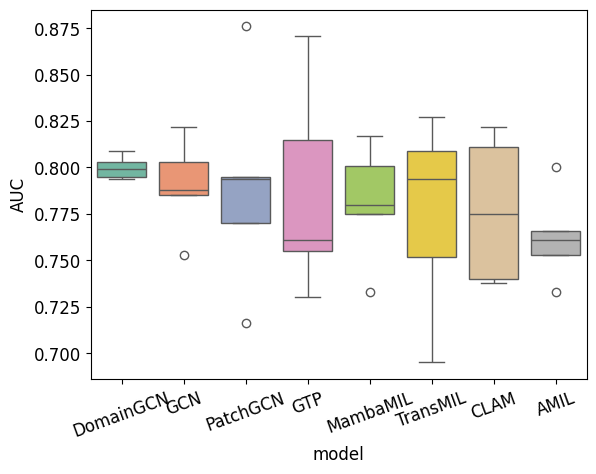}
\caption{Boxplots showing the validation AUC of all models compared in this study. }
 \label{fig:boxplot}
\end{figure}

\subsection{Model training and evaluation}
DomainGCN was trained with a learning rate of 0.0001 with Adam optimizer. The training was stopped early if validation loss did not increase for more than 5 epochs and the maximum number of epochs was 200. Ulcer-label-stratified 5-fold cross-validation (in each fold, 80\% training and 20\% validation) was used to train and evaluate the model. All models' performance was compared using Area Under Curve (AUC), F1 score, and Accuracy (ACC) from validation sets.

\section{Results and Discussion}
Table \ref{table: table1} contains the quantitative results of all models and Figure \ref{fig:boxplot} visualizes the distribution of validation AUC. Results show that graph-based approaches including GTP, PatchGCN, GCN, and DomainGCN generally outperform all MIL-based methods AMIL, TransMIL, CLAM, and MambaMIL across three evaluation metrics. This indicates that the message-passing mechanism of these graph models might have captured useful spatial context, unlike MIL-based methods that do not take into account spatial context. This trend is evident when comparing the high attention regions of DomainGCN with two most popular MIL methods in Figure \ref{fig:visualize}(c)(d)(e), where DomainGCN produced spatially more consistent and plausible attention that mostly reside in the ulcer regions characterized by bleeding, inflammation, and epithelium detachment as compared to AMIL and CLAM, which have more spread-out attention regions that were not ulcers (false positives). Adding domain knowledge into GCN resulted in the best average AUC, F1 score, and ACC and lowest standard deviation across 5 folds compared to all other graph baselines, indicating that the domain knowledge further improved the consistency of model performance. 

To date, very few studies have applied machine learning methods to detect ulcer regions in IBD using H\&E-stained WSIs. Mokhtari \textit{et al.} \cite{mokhtari2023interpretable} developed a weakly-supervised model to predict disease relevant features in IBD. One of their tasks was also binary classification on ulcer labels. They used SOTA models based on MIL and transformers but did not explore graph-based methods. Our study indicates the benefit of using a graph model to identify ulcer regions in IBD. Further, the dataset used by Mokhtari \textit{et al.} \cite{mokhtari2023interpretable} was also from IBD Plexus but they used macroscopic labels, which sometimes might not be consistent with the pathological ground truth. As mentioned in section \ref{dataset}, this study was done on the biopsy samples from the ileum since samples from different locations have different morphology so it might be better to tailor a model to each location. In addition, our method shows better results as compared to Mokhtari et al.'s \cite{mokhtari2023interpretable}. 

To summarize, we proposed a domain-driven GNN named DomainGCN for predicting whether there is an ulcer in a WSI of IBD patients. The domain knowledge was injected to the GNN nodes as importance scores by considering the amount of epithelium, lymphocytes, and debris in each node. The effectiveness of DomainGCN was illustrated by comparing it with multiple SOTA methods. Further work includes improving the generalizability by diversifying training data, exploring the generalizability of DomainGCN in other locations of the digestive tracts, optimizing for the computational cost, and comparing DomainGCN with more SOTA graph models. Eventually immune cells will be detected and analyzed from the DomainGCN-identified ulcer regions, which can help us understand the biology and treatment options of IBD. 

\section{Compliance with ethical standards}
The results published here are in whole from the Study of a Prospective Adult Research Cohort with IBD (SPARC IBD). SPARC IBD is a component of the Crohn’s \& Colitis Foundation’s IBD Plexus data exchange platform. SPARC IBD enrolls patients with an established or new diagnosis of IBD from sites throughout the U.S. and links data collected from the electronic health record and study specific case report forms. Patients also provide blood, stool and biopsy samples at selected times during follow-up. The design and implementation of the SPARC IBD cohort has been previously described.

\bibliographystyle{ieeetr}
\bibliography{refs}

\end{document}